\documentclass
[prl,twocolumn,10pt,tightenlines,letterpaper,runinaddress]{revtex4}
\usepackage{graphicx}
\begin{document}
\title{Reply to Cheng, Harko, and Shabad}
\author{Jes Madsen}
\affiliation{Department of Physics and Astronomy, Aarhus University, DK-8000 \AA rhus C, Denmark}
\date{October 14, 2008}
\maketitle

As demonstrated in \cite{Madsen08} there is a universal relation for the
maximal electric charge of any spherical object with radius 
$R\gg m_{e}^{-1} \approx4\times10^{2}$ fm, 
\begin{equation}
Z_{\infty}(R)\approx2m_{e}R/\alpha=0.71R_{\text{fm}},
\end{equation}
where $Z_{\infty}$ is the maximal charge possible with infinite time available
for electron-positron pair creation in a supercritical field, $m_{e}$ is the
electron mass, $\alpha$ is the fine structure constant, and
$R_{\text{fm}}$ is the radius in fm. For $R\approx m_{e}^{-1}$ there is a
smooth transition (denoted $Z^{\text{Mads}}(R)$ in \cite{Cheng08}) 
from the object-specific core charge without
electrons, $Z_{\text{core}}(R)$, to the universal asymptotic
relation $Z_{\infty}(R)$.
The universal
relation changes for $R\gg25m_{e}^{-1}\approx10^{4}$ fm into $Z(R)\approx
11.2m_{e}^{2}R^{2}=7\times10^{-5}R_{\text{fm}}^{2}$, if only a finite time is
available for pair creation \cite{Madsen08}.
\begin{figure}
[b]
\begin{center}
\includegraphics[
angle=-90.,
width=3.4in
]
{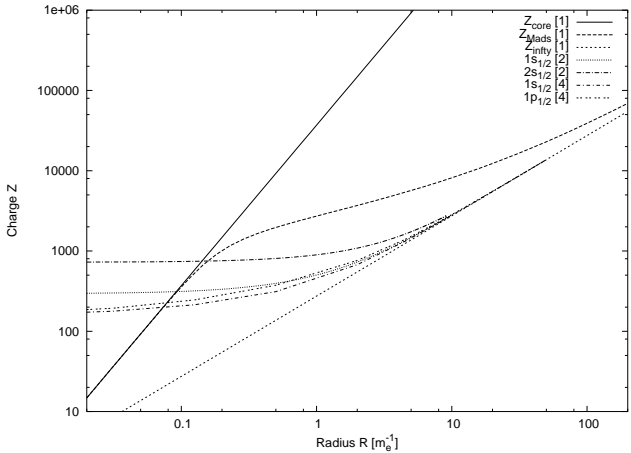}
\caption{$Z(R)$-curves taken from
\cite{Madsen08}, \cite{Cheng08} and \cite{Dicus08}.}
\end{center}
\end{figure}

The authors of \cite{Cheng08} consider only the infinite-time case. They argue
that $Z^{\text{Mads}}(R)$ is only one of two possible boundaries in the
charge-radius plane that act as attractors of charged objects, the other being
the threshold charge $Z_{\text{thr}}^{(1)}(R)$ where the electric charge is
just sufficient for the first possible pair creation to take place. The
charge-radius plane should then be divided into three regimes. 
(i) $Z_{\text{core}
}(R)>Z^{\text{Mads}}(R)$, where pair creation saturates the Pauli principle
with electrons partly shielding the core charge and reducing the net charge of
the sphere to $Z^{\text{Mads}}(R)$ as found in \cite{Madsen08};
(ii) $Z_{\text{thr}}^{(1)}(R)<Z_{\text{core}}(R)<Z^{\text{Mads}}(R)$
\cite{inequality}, where \textquotedblleft the body creates pairs till the net
charge diminishes down to $Z_{\text{thr}}^{(1)}(R)$\textquotedblright
\cite{Cheng08};
(iii) $Z_{\text{thr}}^{(1)}(R)>Z_{\text{core}}(R)$, where no pair creation
takes place. Ref.~\cite{Cheng08} calculates $Z_{\text{thr}}^{(n)}(R)$, where $n$
enumerates s-wave solutions that dive into the negative energy continuum below
$-m_{e}$, by solving the single particle Dirac equation in a large radius
approximation for a spherical shell core potential
and finds that $Z_{\text{thr}}^{(n)}(R)\rightarrow Z_{\infty}(R)$
for $R\rightarrow \infty$. This asymptotic behavior
was previously demonstrated for s- and p-waves for a
shell core charge in \cite{Dicus08} and there taken to confirm the universal
relation of \cite{Madsen08}.

The idea
\cite{Cheng08} that $Z_{\text{thr}}^{(1)}(R)$ should act as an attractor in
the charge-radius plane is based on a misinterpretation of the solutions to
the Dirac equation diving into the negative energy continuum
below $-m_{e}$. In particular, Eqs.~(1) and (2) in \cite{Cheng08} are not
applicable to the self-consistent problem considered in \cite{Madsen08}
since the screening
charge from the electrons formed is not included. A detailed discussion of the
underlying physics is given in \cite{Greiner85}. When $Z_{\text{thr}}
^{(2)}(R)>Z_{\text{core}}(R)>Z_{\text{thr}}^{(1)}(R)$ ($(1)$ and $(2)$
denote the two lowest-lying solutions found in
\cite{Dicus08}, not the s-waves from \cite{Cheng08}) only the 1s$_{1/2}$-state
is diving into the negative energy continuum. This shields the core charge by
two units. For $Z_{\text{thr}}^{(3)}(R)>Z_{\text{core}}(R)>Z_{\text{thr}
}^{(2)}(R)$ additional screening by p$_{1/2}$ electrons becomes available,
etc. As the number of electrons increases, one must
self-consistently include their charge \cite{Greiner85},
\cite{Muller75}. For $Z_{\text{core}}(R)>Z^{\text{Mads}}(R)$ all available
Fermi levels are filled and the relation of \cite{Madsen08} is followed by any
sphere with $Z_{\text{core}}(R)>Z^{\text{Mads}}(R)$. All objects with
$Z^{\text{Mads}}(R)>Z_{\text{core}}(R)>Z_{\text{thr}}^{(1)}(R)$
are partly screened, and a self-consistent analysis could
map out a sequence of $Z(R)$ curves populated in steady state. Contrary to
the claim of \cite{Cheng08}, $Z_{\text{thr}}^{(1)}(R)$ does not act like an
attractor but marks the onset of pair production.

I thank Berndt M\"{u}ller, Constantin Klier, and in particular Johann Rafelski
for useful comments.

\end{document}